**Mapping intra firm trade in the automotive sector: a network approach**

Matthew Smith[1*] & Yasaman Sarabi[1]

[1] Edinburgh Business School, Heriot-Watt University, Edinburgh, UK

[*]Corresponding author, M.Smith_2@hw.ac.uk

**Abstract**

Intra-firm trade describes the trade between affiliated firms and is increasingly important as global production is fragmented. However, statistics and data on global intra-firm trade patterns are widely unavailable. This study proposes a novel multilevel approach combining firm and country level data to construct a set of country intra-firm trade networks for various segments of the automotive production chain. A multilevel network is constructed with a network of international trade at the macro level, a firm ownership network at the micro level and a firm-country affiliation network linking the two, at the meso level. A motif detection approach is used to filter these networks to extract potential intra-firm trade ties between countries, where the motif (or substructure) is two countries linked by trade, each affiliated with a firm, and these two firms linked by ownership. The motif detection is used to extract potential country level intra-firm trade ties. An Exponential Random Graph Model (ERGM) is applied to the country level intra-firm trade networks, one for each segment of the automotive production chain, to inform on the determinants of intra-firm trade at the country level.

## 1. Introduction

The organisation of production has changed in recent decades, with a shift from a product manufactured in its entirety in a single location to a fragmented process, with segments of the manufacturing process occurring in production sites spread out across the globe (Helg and Tajoli 2005; Jones and Kierzkowski 2005; Zhang 2020; Brondino 2023). Driving this fragmentation of production is the change in the organisational form of Multinational

Enterprises (MNEs) which account for the lion's share of economic activity in the global economy (Navaretti et al. 2002; Helpman et al. 2004; Antràs and Yeaple 2014; Helpman 2014). MNEs are increasingly making use of international sourcing strategies in order to coordinate their global supply chain (Helpman, 2006).

The fragmentation of production has subsequently led to an increase in intra-firm trade, that is trade between parent and affiliate (subsidiary) firms (Bonturi and Fukasaku 1993; UNCTAD 2013). The reorganisation and coordination of MNE production processes into geographically segmented production sites has substantially increased the importance of intra-firm trade in the world today (Antràs 2015). The rise and importance of intra-firm trade in the past decades has been recognised by several scholars (Cho 1988; Gilroy 1989; Kim et al. 2023). However, many of these studies rely on data for individual countries (usually developed countries), as statistics and data on intra-firm trading patterns (especially at the firm level and on a global scale) are widely unavailable (Andersson and Fredriksson 2000). Understanding patterns of intra-firm trade has the potential to inform on industrial policy, pertaining to the benefits of trade liberalisation and attracting investment (Lanz and Miroudot 2011).

There are growing concerns among international institutions such as OECD, WTO and UNCTAD that trade and investment statistics collected and published as transactions between nations do not allow sufficient understanding of the increasingly internationally fragmented production processes organised by firms operating across national boundaries (WTO-OECD-UNCTAD 2012). The limited data available on the trading acitivity of large enterprisess greatly constricts effective policy formulation, highlighting the need for the complex nature of production to be reflected in the complexity of datasets under analysis (Feenstra et al. 2010; Fortanier et al. 2020).

The need for improved datasets has been recognised by a number of international organisations. For instance, a collaborative initiative by the OECD – WTO has resulted in the creation of a dataset to map value added based on international input-output tables, the Trade in Value Added (TiVA) dataset (OECD 2013). In more recent years, input-ouput tables have been frequently used to map value added patterns to comment on production trends in the global economy (amongst them Johnson & Noguera, 2012), rather than patterns of intra-firm trade. However, these input-output tables are subject to a number of limitations, firstly they have limited coverage, only available for around sixty nations at the broad macro sector level for 1995 – 2011. Input-output dataset often requiring substantial statistical manipulation and processing before use in empirical studies (McCleery and DePaolis 2014). Nevertheless, they remain highly useful on informing on patterns of value added in the modern global economy.

During its 46th session, the UN Statistical Commission requested a programme of work that would develop and extend the measurement frameworks for international trade and economic globalisation. This session led to the creation of the Inter-Secretariat Working Group on International Trade and Economic Globalisation Studies (ISWG-ITEGS), which aims to promote the development of databases for international trade and economic globalisation studies. In particular, it aims to promote the improvement of micro level statistics for the better calculation of macro level statistics, where there is an emphasis on linking business statistics to international trade data. Additionally, the programme of work outlined by the Statistical Commission aims to develop a global enterprise group register, building on the EuroGroups Register project, utilising a wide range of data sources, including Bureau van Dijk and Dunn & Bradstreet, in order to map the global structure of multinational groups (UNECE 2015).

Smith et al. (2019) recognised this need for an alternative dataset to examine production patterns in the modern global economy, utilising a relational framework to construct a dataset consisting of both firm and country level data. They created a multilevel network of firms linked by ownership at the micro level, countries linked by trade at the macro level and a firm-country affiliation network linking the two. They mapped the patterns of international trade and investment transactions for a single industrial product group, with the product classification used to match the data at various levels. The product perspective has been argued to represent a more relevant framework to explain the fragmentation of production in the modern global economy, as a finished product is now generated through trade in intermediate goods manufactured at various production sites spread out across the globe (Grossman and Rossi-Hansberg 2008; Baldwin and Robert-Nicoud 2014). This paper also draws on a multilevel approach, as it allows to overcome the fictitious separation between the micro and the macro level when examining the production patterns in the global economy, along with operalisationalising the aims of the ISWG-ITEGS initiative, more sepcifically utilising micro level statistics for the better calculation of macro level patterns and trends.

This study aims to contribute to the literature attempting to understand patterns of intra-firm trade by presenting a novel approach, drawing on micro and macro level data to construct a network of (potential) intra-firm trade between countries. The empirical setting for this study is the automotive sector, where intra-firm trade networks are constructed for three segments. These three segments represent a different slice of the global automotive supply chain. Furthermore, the inter-segment intra-firm trade network is examined, that is intra-firm trade patterns between these segments and slices of the automotive production network. Several research questions are addressed regarding the processes that give rise to (potential) country level intra-firm trade ties.

The paper is structured as follows; in the next section an overview of the literature on intra-firm trade is presented, along with the hypotheses and research questions this paper aims to tackle. Following this, a section on the data and methods is presented, providing an overview of the approach and data utilised to construct the intra-firm trade network, along with the methodological approach utilised in this study. In this section, details on the empirical setting are also provided. This is followed by a discussion of the results along with a set of robustness checks. The final section of the paper includes overall concluding comments.

## 2. Literature Review

There have been several theoretical approaches utilised to inform on patterns of intra-firm trade. One of the most widely applied is the property rights theory, as outlined by Grossman and Hart (1986) and expanded by Antras (2003). Property rights theory argues that incomplete contracts govern the trade-off between production decisions of outsourcing and offshoring and that the enforcement of incomplete contracts has a critical effect on trade of intermediate inputs. The property rights approach argues that complex goods and intermediate inputs are more likely to be traded via intra-firm transactions, especially when this involves countries with low levels of contract enforcement, therefore results in the production process retained within the firm boundaries (Linghui 2002; Antràs 2014). Transactions between unrelated parties can require investment in the relationship, as buyers will suffer from time lags in production if they have to source specialised inputs elsewhere during contract disputes, which can become increasingly frequent in an environment characterised by weak contract enforcement (Antràs and Chor 2013).

Within international economics, the gravity model is frequently utilised to explain patterns of international trade (Ward et al. 2013). The gravity model is an econometric model reminiscent of Newton's law of universal gravitation, where the volume of trade between two nations is positively related to their economic size (GDP), and negatively to their distance (Anderson 2011). This model has been applied to a limited set of empirical cases to examine intra-firm trade and patterns of multinational production. There is some evidence that intra-firm trade adheres to the gravity model (Bardhan and Jaffee 2004; Ramondo et al. 2015), yet there is scope to extend this to examine where intra-firm trade adheres to the gravity model in various segments of the global supply chain. This is especially salient given intra-firm trade has been found to vary dramatically across industries (Hanson et al. 2005). There is some debate on the impact of geography on intra-firm trade patterns; Irarrazabal et al. (2013) note that for multinational production patterns geography matters, where affiliate sales fall when distance from the headquarters increases. On the other hand, Bombarda (2013) notes that when exports occur within firm boundaries (intra-firm trade), firms are less sensitive to geographic barriers.

Empirical work informing on patterns of intra-firm trade often relies on detailed national level data (where firm surveys have been conducted, and the results aggregated) (Ruhl 2013; Cho and Choi 2022). Detailed national level data is available for a small range of nations, often focusing on the cases of Japan (Kiyota et al. 2020; Matsuura et al. 2023; Okubo and Watabe 2023) and the USA (Atalay et al. 2014; Ramondo et al. 2016). These dataset often only provide sectoral coverage at broad aggregate level. Nevertheless, studies drawing on national level data provides salient insights on the drivers of intra-firm trade in the modern global economy (Bernard et al. 2010; Ruhl 2015). Chun et al. (2017) provide an empirical analysis of intra-firm trade in Korea and Japan drawing on surveys of business activity data for South Korea and Japan. They confirm the work of Helpman et al. (2004), noting that trade is skewed with firm trade is concentrated in a small handful of large firms or multinational enterprises. Corcos et al. (2013) provide an analysis of the determinants of intra-firm trade for French firms; they note that  intra-firm trade is more likely with countries that have a good judicial systems, and that final good and complex intermediate goods are more likely to occur within firm boundaries (and therefore more likely to be part of intra-firm trade) Further insights have been established on the firm level characteristics of entities that engage in exporting, importing and intra-firm trading activities; that these firms tend to be larger, more productive and more capital- and skill intensive firms (Antràs 2005; Antràs and Rossi-Hansberg 2009; Castellani et al. 2010).

Ivarsson and Johnsson (2000) explore the link between motivations for FDI (in terms of resource, market, efficiency and strategic asset seeking FDI) and the potential for intra-firm trade. In their analysis of firms in Sweden, they identify that efficiency-seeking motives for FDI is associated with intra-firm exports of finished goods and material inputs, whilst market-seeking motives of FDI are associated with intra-firm imports of complementary finished products. They note that resource and strategic assets seeking motives for FDI are negatively associated with intra-firm trade patterns (Ivarsson and Jonsson 2003).

In addition to FDI and intra-firm trade, many have examined the impact of MNEs and their affiliates on the economy, a key policy issue. Vinod and Rao (2019) note that US MNE trading activities, including intra-firm trade, have a positive impact on the global economy. Other note that the benefits from intra-firm trade are similar to those that can be reaped from economic liberation and the participation in global value chains (Miroudot and Ye 2020).

This paper seeks to address the following research questions through the use of a novel approach to construct intra-firm network data:

1. Does intra-firm trade follow the gravity model in the automotive sector? Are countries that are closer, share a land border and share a language more likely to be linked by intra-firm trade? Do larger and more affluent countries gravitate towards each other?
2. Does the property rights theory explain patterns of intra-firm trade in the automotive sector?
3. Are the patterns and determinants of intra-firm trade consistent across various segments (and between segments) of the automotive production chain?

## 4. Data & Methods

In this study, a multilevel network approach is taken, where data at the firm level and country level is used to extract a network of potential intra-firm trade links between countries.

Network analysis has been frequently utilised to inform on production patterns (Fagiolo 2016; Gorgoni et al. 2018b); ranging from the analysis of international trade networks (Krings et al. 2014; Piccardi and Tajoli 2015), value added networks (Cerina et al. 2015; Amador and Cabral 2017; Zádor et al. 2022) and the examination of the product space (Hidalgo et al. 2007; Kali et al. 2013).

The multilevel approach consisted of constructing a set of multilevel networks, following the approach outlined by Smith et al. (2019). The multilevel network consists of a network of firms linked by ownership at the micro level, countries linked by trade at the macro level, and a firm-country affiliation (or investment) network linking the two levels (also referred to as the meso level). The elements of the multilevel network are present in fig 1, a dummy representation of the multilevel network and the three levels separately. The firm and country level data were matched using the product code. Firms at the micro level are selected on the basis of whether they are affiliated with a particular product code, which reflects the firm's activities and production of a particular good. At the macro level, the country level network is international trade in this particular good: trade in goods affiliated with this product code. The international trade data is taken from UN Comtrade, and the firm information (including ownership and investment) is extracted from Bureau van Dijk's Orbis.

***Figure 1 Dummy Multilevel Network***

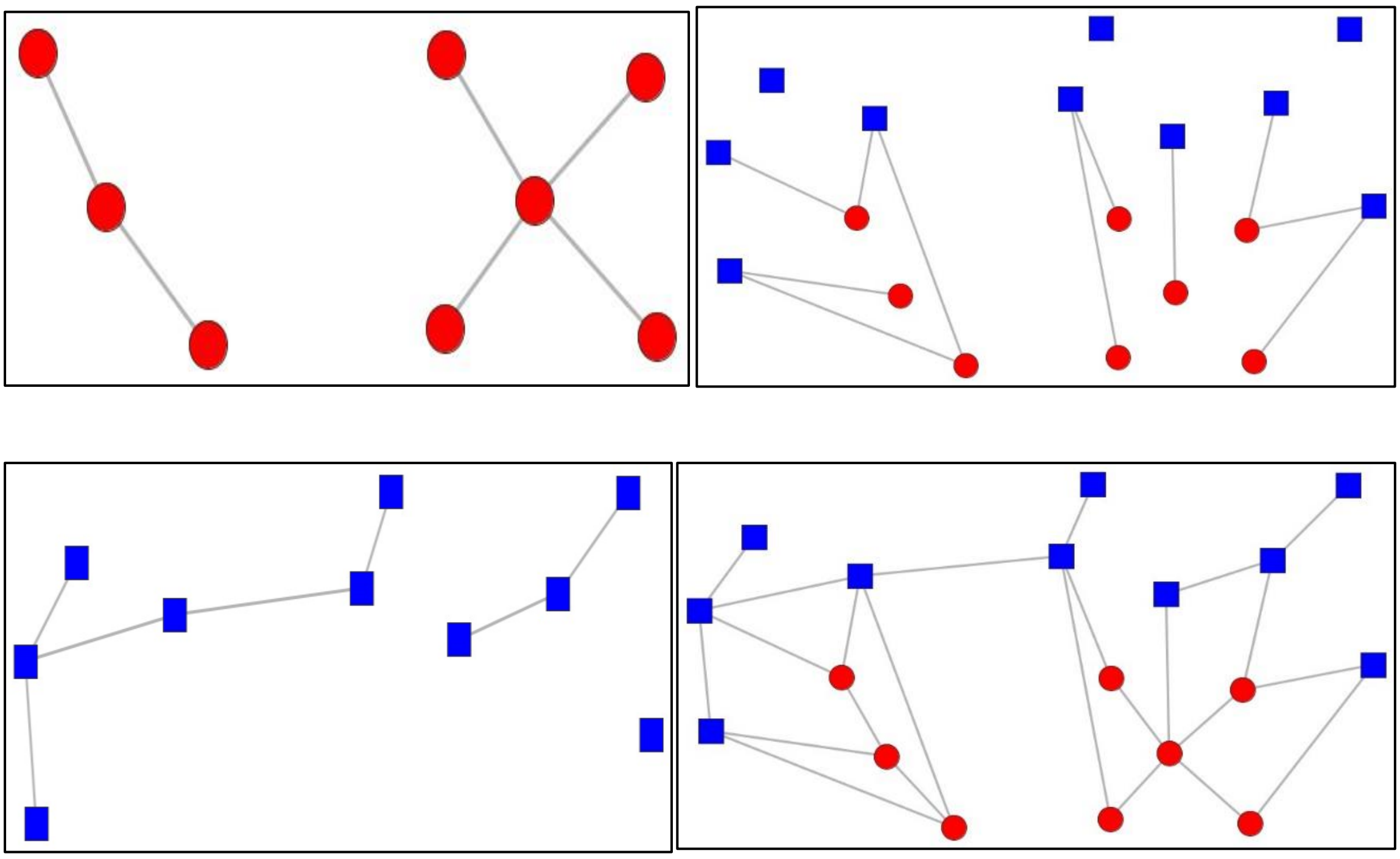

*Note: Top left – ownership network (micro level), Top right – country-firm affiliation network (meso level), Bottom left – trade network, Bottom right – full multilevel network. Red circles – firms and blue squares – countries.*

Smith et al. (2019) make use of a multilevel Exponential Random Graph Model (ERGM) to analyse a multilevel network of international trade and investment. An ERGM is a complex network model; variables are specified in terms of local network structures, where the estimate values from the model indicate the weight and direction of local network configurations in explaining the overall network structure (Lusher et al. 2013). In the case of the multilevel network, local network configurations involve actors from various levels, both firms from the micro level and countries from the macro level (Wang et al. 2013). An example of a network configuration, utilised by Smith et al. (2019) is presented in fig 2; they used it to refer to intra-firm trade. The firms are linked by ownership and are linked to countries that also trade. In the model specification this configuration was specified to examine whether the specific sector under analysis was characterised by intra-firm trade, whether this configuration contributed significantly in explaining the global structure of the multilevel trade and investment dataset.

In this paper, a motif or pattern detection approach is utilised to extract the intra-firm trade configuration in the multilevel network, as shown in fig 2. This information is then used to construct a network of intra-firm trade amongst countries, where the original trade data is filtered only retaining the trade ties found in the intra-firm trade motif detected in the multilevel trade and investment network. A motif or pattern detection approach refers to when a network pattern is detected, examined or extracted from the entire network. This approach has been used in a number of empirical settings (Takes et al. 2018; Liu et al. 2018), including the international trade network (Maratea et al. 2016). Squartini and Garlaschelli (2012) employ a motif detection approach to the international trade network, more specifically to analyse triadic motifs and what explains the formation of these motifs or patterns. Cingolani et al. (2015) detect modular structure and subnetworks in the international trade network in order to test whether there was a tendency to concentrate trade on one or few partners.

***Figure 2 Intra-firm trade configuration***

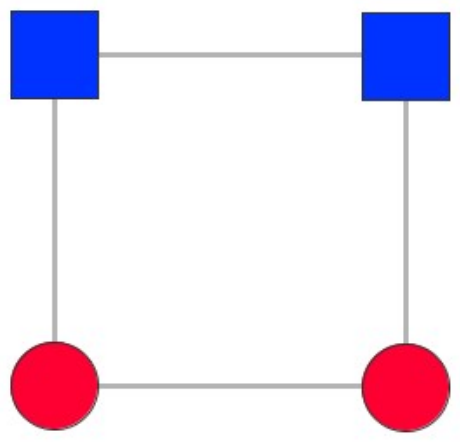

Only the case demonstrated in fig 2 is considered, where there are only four edges linking these four nodes are considered. If a firm is affiliated with both countries, then this may not truly represent intra-firm trade, and therefore is excluded to ensure that intra-firm trade is not over estimated.

It is important to note that in the construction of this filtered network representing potential intra-firm trade transactions between countries, a number of assumptions are made. We assume that firms embedded in these configurations point towards intra-firm trade and assume there is intra-firm trade amongst these countries. Therefore, although in this paper we refer to these networks as intra-firm trade networks, they only indicate the potential for intra-firm trade, as we may potentially capture trade ties that should not be characterised as intra-firm.

As a robustness check for the analysis, we examine whether the structural configurations in the multilevel network (visualised in fig 2, representing intra-firm trade), are a significant feature of the network, and are more likely to occur than in a set of similar randomised networks. We draw on the Z-Score approach outlined by Milo et al. (2002) to evaluate the importance of the intra-firm trade motif (as shown in fig 2) in the observed multilevel network, compared to a set of random networks. For each observed multilevel network, we simulate 1000 random networks (under the Erdős and Rényi 1959 process), with the number of nodes and edges fixed in the simulations. The Z-Score is calculated as follows:

$$Z = \frac{f_{real} - average(\overline{f_{rand}})}{std(\overline{f_{rand}})}$$

The Z-Score is calculated for each multilevel network, where $f_{real}$ is the number of times the intra-firm trade configuration (as visualised in fig 2) appears in the observed multilevel network. $\overline{f_{rand}}$ is the number of times that the intra-firm trade configuration appears in the randomised or simulated networks. In the Z-Score formulation, the average and standard deviation of $\overline{f_{rand}}$ is used. The results of this robustness check as presented in section 6 of the paper.

The empirical setting for the analysis is the automotive sector, where the intra-firm trade networks are constructed for various segments of the automotive global value chains. The automotive sector is an important sector, with large political value, as it is highly visible and contributes significantly to national employment (Sturgeon et al. 2008; Sturgeon and Van Biesebroeck 2011). The automotive sector is a contract-intensive sector (Blanas and Seric 2018), therefore there is the potential for intra-firm trading strategies to be utilised by firms to overcome these issues.

As noted in the introduction, the fragmentation of production has resulted in the production process no longer taking place in a single location, rather split into segments or slices, spread out across the globe into Global Value Chinas (GVCs) (Gereffi et al. 2005). This has resulted in the task or segment level becoming the more appropriate level of analysis (Grossman and Rossi-Hansberg 2008). Therefore in this study an examination of the networks of various segments of the automotive sector is provided; following the approach of Gorgoni et al. (2018a), Smith et al. (2016) and Amighini and Gorgoni (2014), the segments include electrical parts, engines and rubber and metal. These segments are characterised by varying level of technological content, ranging from high tech electrical parts to low tech rubber and metal components, therefore the organisation of production of each segment will vary (Smith and Sarabi 2020). The product codes for each segment are outlined in the appendix, trade and firms associated with the codes are examined for each network. For these three groups, when constructing the intra-firm network, the firms in fig 2 belong to the same sector classification; for example in the electrical parts network, only the firms operating in the production of automotive electrical parts and the ownership ties between them are considered.

A further network is created examining firms in all segments, the ties between firms belonging to different segments and trade in all the segment product groups, in order to inform on potential intra-firm trade between segments. Therefore, in this case only the ownership ties between firms operating in different segments are considered in the motif presented in fig 2 to create the inter segment intra-firm trade network (contrasting to the individual segment networks).

The trade networks produced from the pattern detection of the multilevel networks are then analysed in order to address research questions pertaining to what gives rise to the formation of intra-firm trade at the country level.

Many approaches modelling pattern of international trade assume independence, ignoring multilateral dependencies. With products no longer being manufactured in a single location and increasingly made of parts produced in multiple nations, the independence assumption becomes too strong (Ward et al. 2013), especially for intra-firm trade. Therefore, there is need to overcome the issue of assumed independence required for the majority of regression models. In this paper, we make use of Exponential Random Graph Model (ERGM) to an undirected one-mode network of intra-firm trade, in order to explain the process unpinning the formation of these trade ties.

The ERGM takes the following form:

$$P(Y = y) = \frac{1}{k(\theta)} \exp(\sum \theta_Q z_Q(y))$$

Where:

$Y$ is the observed network

$y$ is a network instance

$Q$ is all the network configuration types (the local structural configurations)

$z_Q(y)$ is the network statistic corresponding to configuration type $Q$.

$\theta_Q$ is the parameter corresponding to configuration type $Q$.

$k(\theta)$ is the normalizing constant to ensure that the above is a proper probability distribution.

ERGMs have been applied in a wide variety of empirical settings, including the analysis of international trade and investment patterns (Smith et al. 2019; Herman 2022; Smith and Sarabi 2022).

In the model, we specify structural terms, along with country level attribute terms. The structural terms include an edges term; this captures the baseline tendency for ties to form in the network (the edges term is analogous to the intercept term in a regression analysis). A centralisation parameter is specified, the geometrically weighted degree centrality parameter. A positive and significant effect would indicate that the trade ties are distributed evenly across the network, whereas a negative and significant effect would suggest that trade ties are concentrated in a small handful of countries Further terms include clustering terms: geometrically weighted edgewise shared partner (GWESP) and geometrically weighted dyadic shared partner (GWDSP) (Snijders et al. 2006; Goodreau 2007; Hunter 2007). GWESP captures transitivity in the network, the tendency for connected actors to have multiple shared partners. GWDSP captures the propensity for actors (not necessarily connected) to have multiple third shared partners. Therefore, when interpreted together, they can inform on the tendency for triadic closure (Leifeld and Schneider 2012). For instance, a positive and significant GWESP, and negative and significant GWDSP would indicate, there is a tendency for open triads to close in the network.

We test whether a number of country level attributes explain the presence of a potential intra-firm trade tie between countries. More specifically, we test whether intra-firm trade is aligned with the gravity model of international trade and whether intra-firm trade is explained by the property rights theory. To test whether the formation of (potential) intra-firm trade ties is aligned with the gravity model of intra-firm trade, several variables are included: GDP (market size), GDP per capita (market affluence), distance, whether countries share a border and whether countries share a common language. For GDP and GDP per capita, two variables are specified, an activity and difference variable. The activity effect captures whether countries with a larger attribute value are more likely to form trade ties. The difference attribute captures likelihood of forming a tie as a function of the absolute difference between the attribute value of the countries.

In order to test the property rights theory, two variables are used to capture this, extracted from the World Bank, following the approach utilised by Blanas and Seric (2018). Firstly, the number of days for the enforcement of contracts; a shorter time refers to better contract enforcement. Secondly, the rule of law index, where higher values indicate a stronger rule of law in these countries. For these parameters, an activity and difference effect are also specified.

In the model specification, the number of firms operating in the sector in a country is also examined and how this interacts with the propensity for intra-firm trade. This information is taken from the original multilevel network, the number of firms in the segment group operating in the country. An activity and difference term are included for this investment data. This allows to examine whether intra-firm trade is deterred or encouraged by there being a large number of relevant firms present in the country.

## 5. Results

Firstly, an overview of the firms in each segment is provided, the descriptive statistics of the full set of firms listed on Orbis operating in the segment, then an overview of the firms in the connected component, that is firms linked to other firms by ownership. These firms drive the filtering process in the motif detection and extraction of the potential intra-firm trade ties.

*Table 1 Firm Descriptive Statistics*

| | All Firms | | | Connected Ownership Network | | |
|---|---|---|---|---|---|---|
| | **Electrical Parts** | **Engines** | **Rubber & Metal** | **Electrical Parts** | **Engines** | **Rubber & Metal** |
| **Number of Firms** | 887 | 5418 | 1299 | 109 | 709 | 120 |
| **Operating Revenue Mean** | 572972.3 | 564887.3 | 522760.3 | 1395599.4 | 1467822 | 2118332.7 |
| **Operating Revenue SD** | 3285468 | 3220951.7 | 2648009.7 | 7730942.3 | 6671464 | 6638576 |
| **Number of Employees Mean** | 2775.6 | 2456.1 | 2387.7 | 4619.4 | 6032.6 | 10952.7 |
| **Number of Employees SD** | 9140 | 11157.1 | 11560.6 | 13320.5 | 20776.4 | 34534.8 |

In the various segments, a smaller proportion of firms are in the connected components, and these firms are on average larger (in terms of number of employees and operating revenue). This provides some support for the use of these firms in constructing the country level intra-firm network, as extant work on trading firms suggests that they tend to be larger and more productive, with trade concentrated in a smaller set of firms (Bernard et al. 2007; Haller 2012).

Fig 4 – 7 present the visualisations of the intra-firm trade networks for the various segments and inter segments networks; the size of the countries and labels refers to the number of trade connections in the network. Table 2 presents further descriptive network statistics for these networks (an overview of these measures is provided by Borgatti et al. 2018). Density refers to the level of connectivity in the network. Average degree is the average number of trade ties each country in the intra-firm network has. Table 2 indicates that in the case of engines and rubber and metal parts, the intra-firm trade network has a higher number of countries participating and denser networks with a higher average degree, compared to the electrical parts and inter segment transaction networks. This suggests that intra-firm trade is concentrated in a smaller handful of actors in the case of electrical part and inter segments within the automotive supply chain. Degree centralisation captures the spread of these degree centrality scores in a network; in this case it is used to examine the distribution of intra-firm trade ties. A high degree centralisation indicates that trade ties are concentrated in a handful of countries in the network, whilst a low score would suggest that they are spread evenly throughout the network. In the case of engine and rubber and metal, the intra-firm trade ties are concentrated in a handful of countries, especially compared to the inter segment network.

Degree assortativity captures the correlation between degree centrality of connected actors in a network (Newman 2002), where scores can be between -1 and 1. A positive score would suggest that well-connected countries with a high degree score link to other well-connected countries. Whereas a negative score, which would suggest that high degree countries are connected to many low degree countries, pointing towards a pattern of disassortativity in the network. In table 2, in all component groups the score is negative, pointing towards patterns of disassortativity, with countries with a high number of trade ties linked to many peripheral, low degree countries.

***Table 2 Network descriptive statistics***

| | **Electrical Parts** | **Engines** | **Rubber & Metal Parts** | **Inter Segment** |
|---|---|---|---|---|
| **Size** | 15 | 39 | 32 | 17 |
| **Density** | 0.1524 | 0.2038 | 0.2016 | 0.1471 |
| **Average Degree** | 2.1333 | 7.7436 | 6.25 | 2.3529 |
| **Degree Centralisation** | 0.2762 | 0.3752 | 0.4435 | 0.1654 |
| **Degree Assortativity** | -0.4803 | -0.243 | -0.2216 | -0.2408 |

International trade network are often characterised by a core-periphery strucutre (Smith and Sarabi 2022), therefore, we apply a core-periphery analysis (as described by Borgatti and Everett 2000) to identify which countries are part of the connected core of the trade network, and what countries play a more peripheral role in intra-firm trade. Fig 3 presents the core – periphery analysis for each component group, where countries belonging to the core are highlighted in dark blue, whilst those in the periphery are highlighted by light blue. The countries in grey are no present in the intra-firm trade network.

***Figure 3 Core-Periphery Analysis***

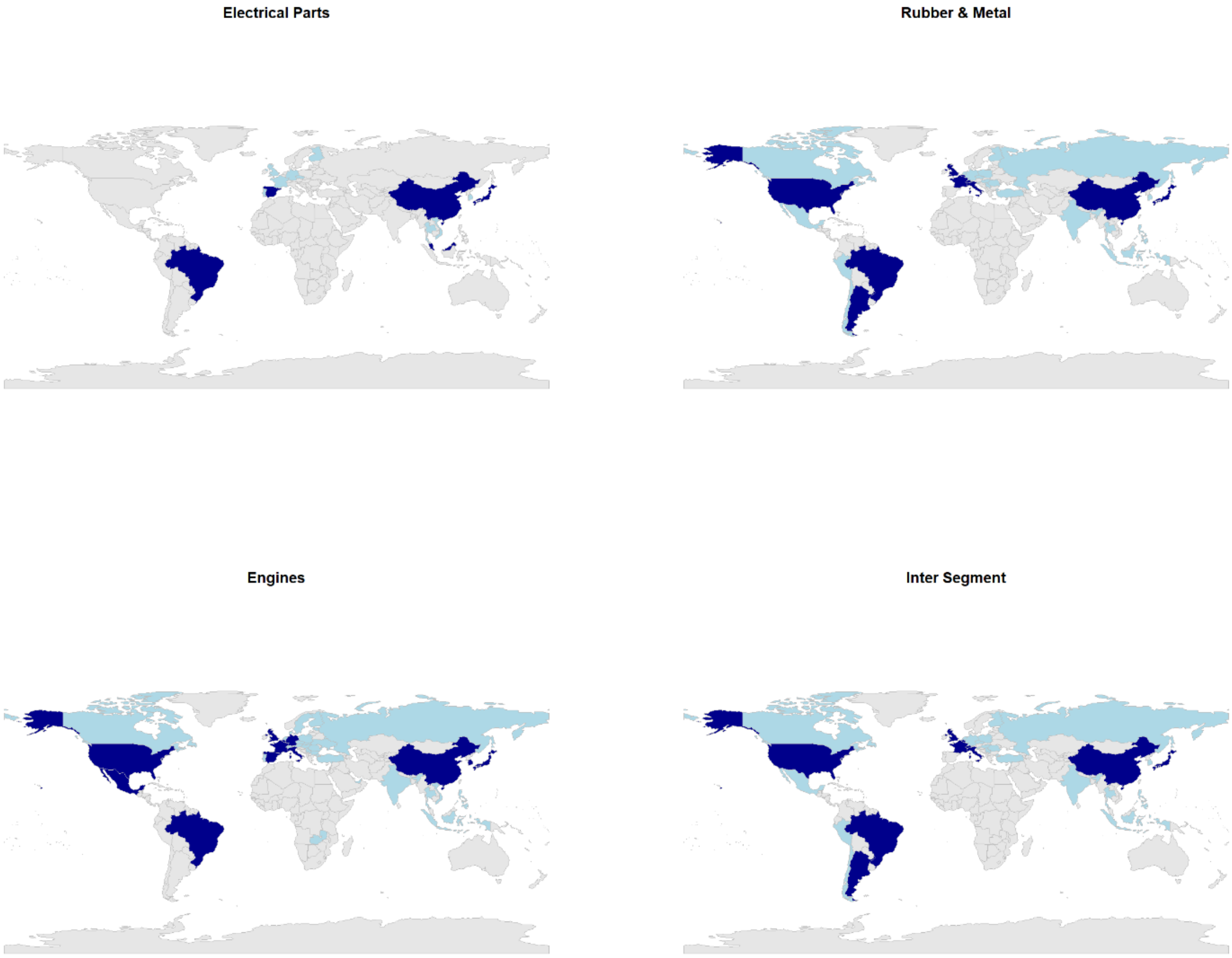

Fig 4 presents the filtered country level network of (potential) intra-firm ties in the electrical parts segment of the automotive production process. We observe that there are two separate components, a European component and a chiefly Southeast Asian and Pacific component (with a handful of exceptions). It is not surprising that there is a high number of countries from Southeast Asia and Pacific, given the competitive advantages of the region in the wider electronics GVC (Sturgeon and Kawakami 2010). The European section of the network is centred on Spain; Spain has developed a favourable environment for the production of automotive components (Lampón et al. 2015, 2016). In fig 3, we also observe that Spain is one of the few countries classified as belonging to the core of the network.

*Figure 4 Electrical parts filtered trade network*

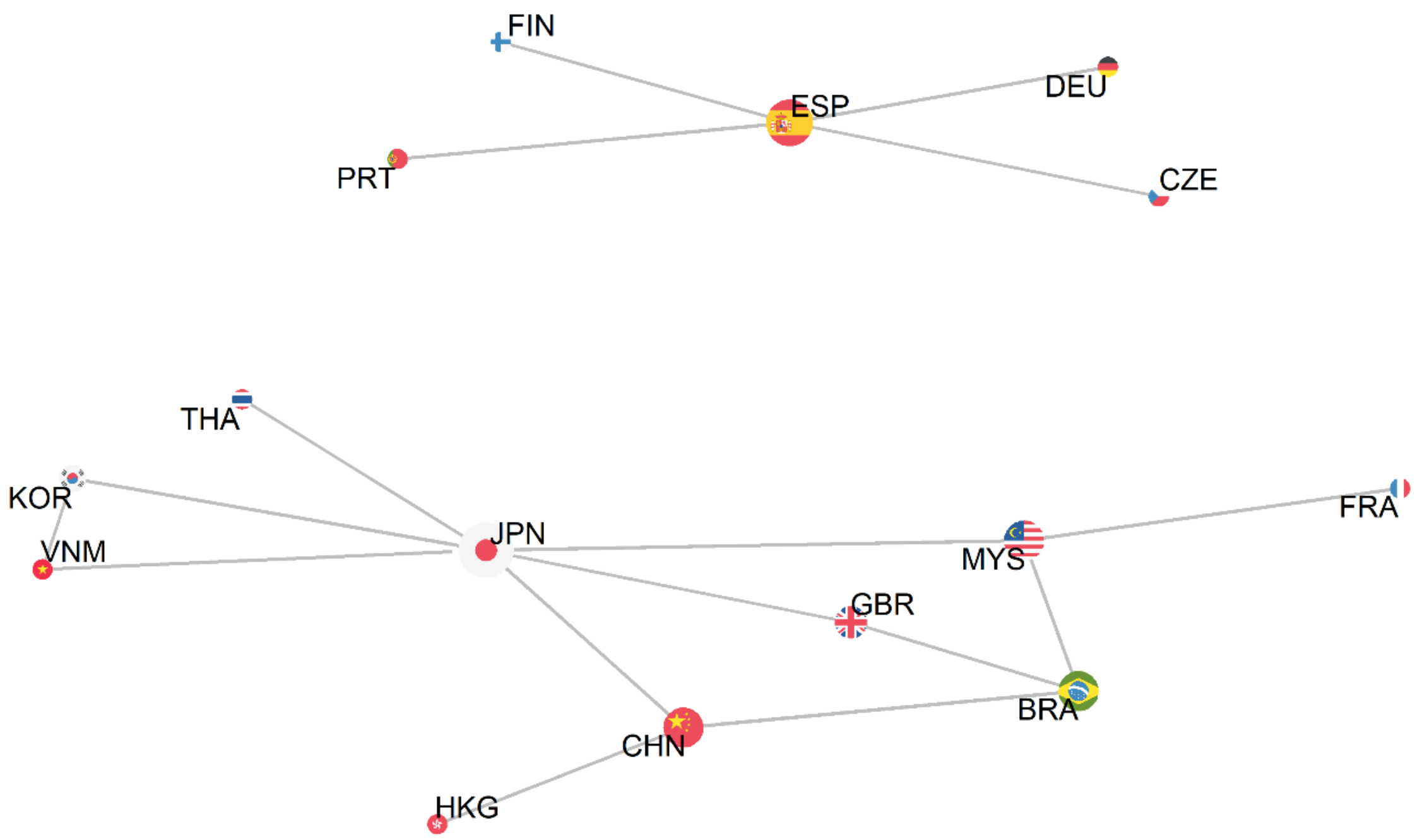

Fig 5 presents the filtered intra-firm trade network for trade within the engines segment, a high to medium tech segment of the automotive production chain. Here we observe that the traditional core players in the automotive sector are at the centre of the network, such as France, Germany and Japan (this can also be observed in fig 3). Towards the periphery of the network, as observed in fig 3, there are a number of nations from emerging economies, along with Eastern European nations. Eastern European nations have expanded capabilities in the automotive sector in recent years (Cieślik et al. 2019; Pavlínek 2020) is reflected here. Similar to electrical parts, Brazil is also a key player in this network (belonging to the core of the network); this indicates that in the automotive production process, firms select to retain production within the firm boundaries and participate in intra-firm trade. What can be observed from fig 4 and 5 is that European nations, and East Asian & South Pacific players are more central than the US in terms of intra-firm trade in the automotive sector. Yet the USA is often the empirical setting to understand intra-firm trade and is a traditional dominant player in the automotive sector.

*Figure 5 Engines filtered trade network*

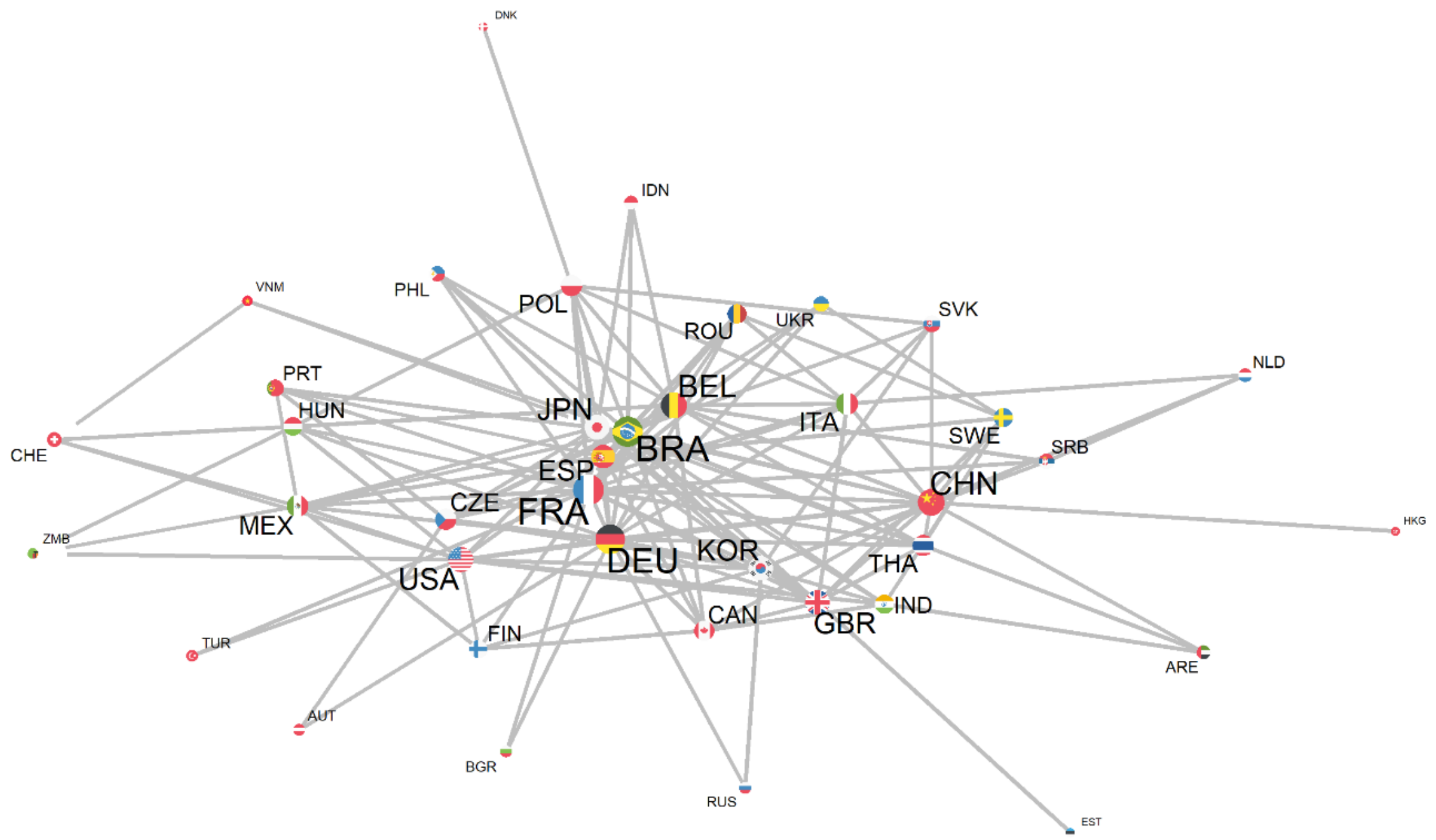

Fig 6 maps potential intra-firm trade in rubber and metal parts. The South American nations of Brazil and Argentina hold key positions in the network, again similar to the other segments of the automotive production chain. France and Italy appear to be the key European nations in the rubber and metal intra-firm network; in fig 3 we can observe that both of these nations belong to the core of the network. In fig 6, there appears to be an areas of the network centred on the USA, and Japan.

*Figure 6 Rubber & metal parts filtered trade network*

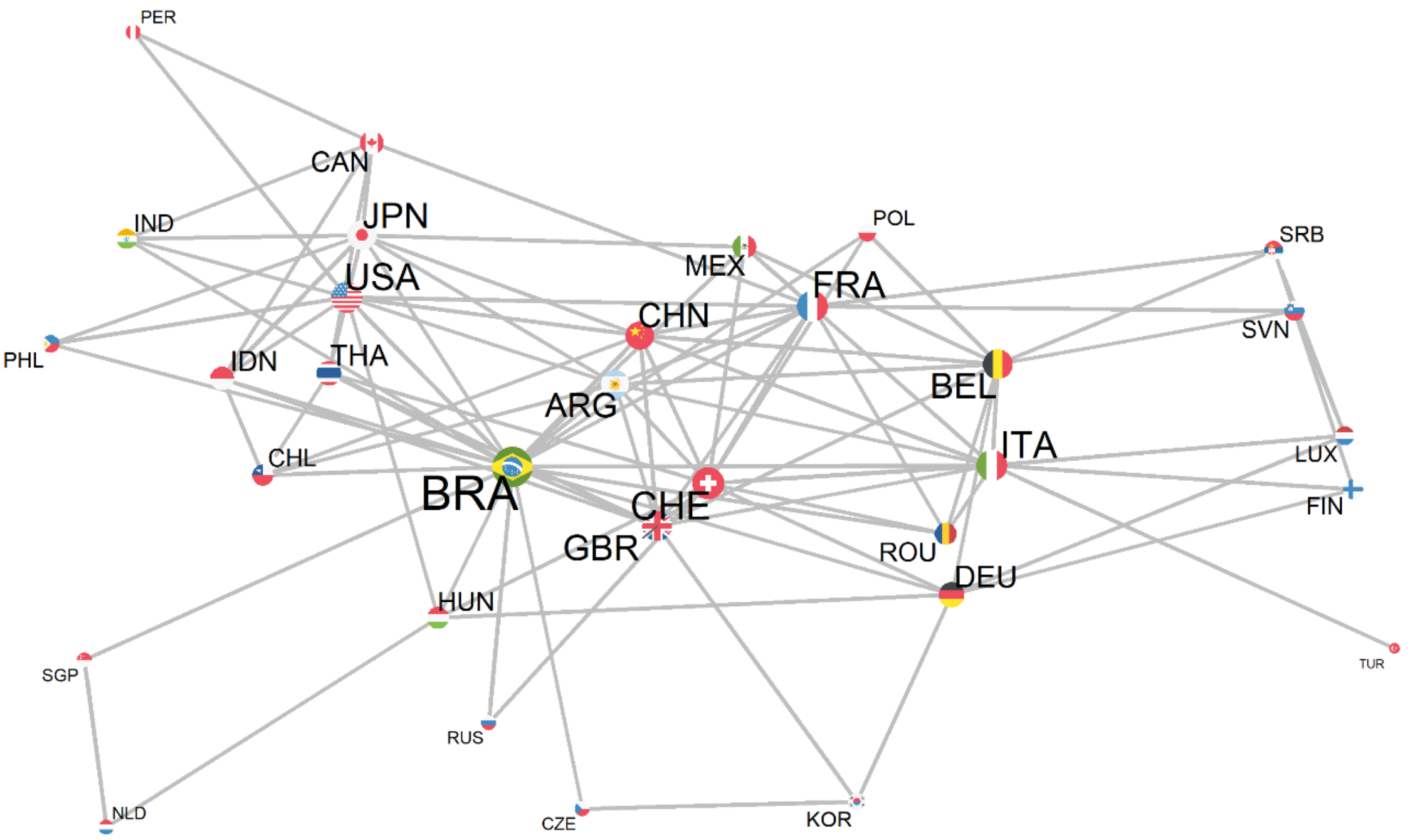

Fig 7 refers to intra-firm between segments of the automotive production network, capturing potential trade ties between firms operating in the three different product groups. Unsurprising there is a separate component for Australia and New Zealand, reflecting their geographic proximity. Many of the traditional players (USA, France, Germany), and key emerging economies, such as China (Amighini 2012), are notable actors in the inter segment network. In fig 3, we observe these nations at the core of the network. Similar to the other segments, Brazil holds a prominent position; this suggests intra-firm trade should be a key component of the country's industrial policy, and there should be efforts by national organisations to develop a Brazilian intra-firm dataset.

***Figure 7 Inter segment filtered trade network***

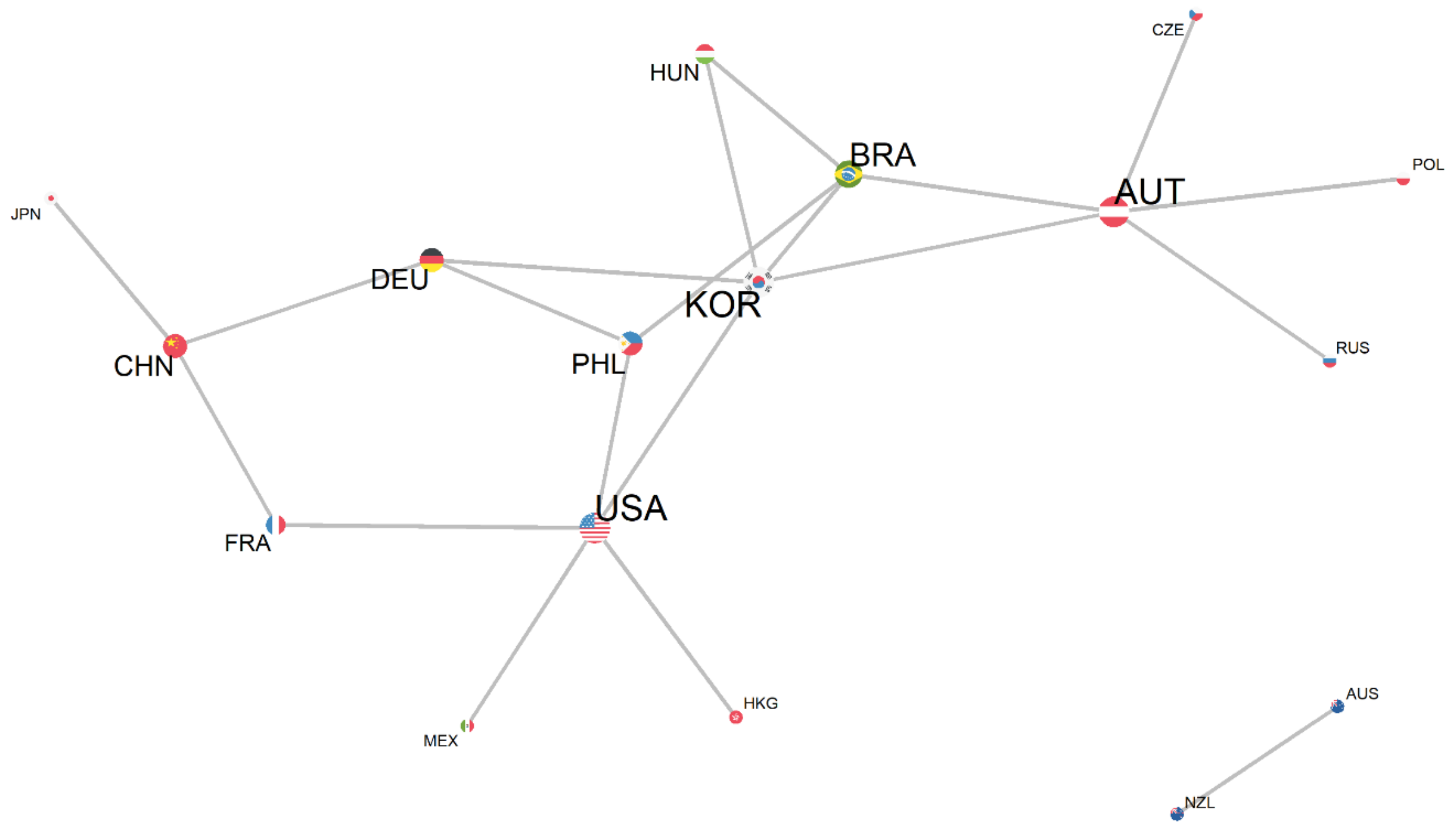

Table 3 presents the results of the ERGMs for the various product groups; these are estimated using the *ergm* package in R (Hunter et al. 2008b). The goodness of fit output is provided in the appendix; the plots indicate that the ERGM models are able to replicate the salient features of the observed networks fairly well (Hunter et al. 2008a). The results highlight differences that emerge between component groups, for both country attributes and structural effects. This is in line with extant work on intra-firm trade, which highlights that differences will emerge between industries, in this case different segments with different levels of technological content.

*Table 3 ERGM Results*

| | Electrical Parts | Engines | Rubber & Metal Parts | Inter Segment |
|---|---|---|---|---|
| Edges | 3.2163*** | -5.0862*** | -2.6465*** | -5.8992*** |
| | (0.1081) | (0.0152) | (0.0142) | (0.0549) |
| Degree | 4.0052*** | -0.1623*** | -1.1322*** | 1.0662*** |
| | (0.0455) | (0.0196) | (0.0167) | (0.0360) |
| GWESP | -0.3763*** | 0.3970*** | -0.0942*** | -0.0861 |
| | (0.0854) | (0.0052) | (0.0203) | (0.0889) |
| GWDSP | 0.3124 | 0.1278*** | 0.0698 | 0.2521 |
| | (0.2019) | (0.0214) | (0.0577) | (0.1597) |
| GDP Activity | 1.9670*** | 0.1009*** | 0.8731*** | -0.6595*** |
| | (0.1278) | (0.0297) | (0.0383) | (0.0754) |
| GDP Difference | 1.1770*** | 0.4729*** | 0.4027*** | 1.4151*** |
| | (0.1594) | (0.0537) | (0.0529) | (0.1256) |
| GDP per capita Activity | -0.1519 | -0.1597*** | -0.8162*** | -0.1508 |
| | (0.1290) | (0.0290) | (0.0338) | (0.0780) |
| GDP per capita Difference | -1.7738*** | -0.2872*** | -0.3221*** | -0.0274 |
| | (0.1433) | (0.0470) | (0.0489) | (0.1213) |
| Rule of Law Activity | -3.4033*** | 0.0784** | -0.0500 | 0.5709*** |
| | (0.1101) | (0.0273) | (0.0449) | (0.1043) |
| Rule of Law Difference | 0.5097*** | 0.0659 | 0.1837*** | -0.8019*** |
| | (0.1231) | (0.0475) | (0.0481) | (0.1252) |
| Contract Enforcement Activity | -0.0023* | 0.0009*** | 0.0002 | 0.0008 |
| | (0.0010) | (0.0002) | (0.0003) | (0.0008) |
| Contract Enforcement Difference | -0.0002 | -0.0016** | -0.0001 | -0.0027 |
| | (0.0035) | (0.0005) | (0.0006) | (0.0021) |
| Investment Activity | 0.0857* | 0.0113*** | 0.0183*** | 0.0023 |
| | (0.0364) | (0.0013) | (0.0043) | (0.0013) |
| Investment Difference | -0.0894* | -0.0112*** | -0.0173*** | -0.0025 |
| | (0.0372) | (0.0013) | (0.0046) | (0.0014) |
| Distance | -0.0002 | -0.0000 | 0.0000 | -0.0000 |
| | (0.0001) | (0.0000) | (0.0000) | (0.0000) |
| Common Language | 1.1628*** | 0.5235*** | 0.3856*** | 1.3416*** |
| | (0.0253) | (0.0087) | (0.0084) | (0.0175) |
| Shared Border | 0.5524*** | 0.0427*** | 0.2503*** | 0.1073*** |
| | (0.0182) | (0.0087) | (0.0088) | (0.0163) |
| AIC | 100.3417 | 601.6830 | 480.3757 | 135.2459 |
| BIC | 145.4591 | 680.0190 | 551.8874 | 184.7611 |
| Log Likelihood | -33.1709 | -283.8415 | -223.1878 | -50.6230 |

***p < 0.001, **p < 0.01, *p < 0.05

For electrical parts and inter segment groups, there is a positive and significant degree parameter; this indicates that intra-firm trade ties are spread evenly amongst the countries within the network. Whereas in rubber and metal parts, these trade ties are concentrated in a small handful of nations. This is consistent with the network maps presented in fig 4 – 7, where in the case of rubber and metal parts there are dominant countries with a high number of intra-firm trade ties. The electrical components and inter segment groups are more likely to be characterised by higher level of complexity in the production process and intra-firm transactions.

The GWESP parameter is negative and significant in the case of electrical and rubber and metal parts, this indicates that there is a tendency against transitivity. In the engines group, the GWESP is positive and significant, indicating that there is a tendency for transitivity in the network, the tendency for connected countries to have multiple shared trading partners. This indicates that clustering is a feature of the electrical and rubber and metal part network, this suggests that clustering is not necessarily linked to the technological content of the product group. GWDSP is only significant in the engines groups, where the results indicate that there is the propensity for countries to share multiple trading partners. When examining the GWESP and GWDSP results together, we observe that there is no significant tendency for open triads to close, in these intra-firm networks.

The GDP effects allow us to examine whether intra-firm trade adheres to the gravity model, examining the interplay between market size and the propensity to form intra-firm trade ties. The activity results indicate that for the three main component groups, there is a tendency for larger nations to form intra-firm trade ties. This provides some evidence that intra-firm trade patterns follow the gravity model in these cases. Whereas for the inter segment group, the result is negative and significant pointing towards smaller nations more likely to form inter segment intra-firm trade ties. The GDP difference effect is positive and significant across all product groups, this indicates that a tie is more likely to be observed between nations with a large difference in market size. This does not necessarily support the gravity model hypothesis that larger nations gravitate toward each other. Rather the difference term results, in combination with the activity result, suggests that there is a tendency for large nations to form intra-firm trade ties with many smaller nations.

The GDP per capita parameter allow us to explore the link between market affluence and intra-firm trade within the automotive sector. The activity term is only significant in the engine and rubber and metal segments, where the negative result here suggests less affluent nations are more likely to establish intra-firm trade ties in these segments. In terms of the difference parameter, this is negative and significant across the three main product groups, yet non-significant for the inter segment group. This indicates that intra-firm trade ties are more likely to occur amongst countries with similar levels of market affluence.

Additionally, parameters specified to examine whether intra-firm trade adhered to the gravity model in the automotive sector included distance, common language and shared border. Distance is not observed to have a significant impact on the formation of these intra-firm trade ties, contrasting to the prediction of the gravity model. However, shared border is positive and significant across all product groups, indicating that distance matters at proximate levels, and that intra-firm trade ties are more likely to occur with neighbouring countries in the automotive sector, perhaps pointing towards regional organisation of production. Common language has a positive and significant effect across the product groups, indicating that intra-firm trade is more likely between countries that share a language. This provides some support for the gravity model. According to property rights theory, contract enforcement is a key aspect of intra-firm trade; Rauch (1999) argues that common language and colonial ties are more important for trade when search and contract-enforcement costs are higher. The positive and significant common language effect also confirms previous work, noting that common language is particularly important for complex differentiated products and trading transactions (such as intra-firm trade).

Additional parameters that were included to test the property rights theory were the rule of law and contract enforcement parameters. In the case of electrical parts, the rule of law activity term is negative and significant, indicating that in this segment, countries with a weaker rule of law are more active in intra-firm trade ties. Whereas for rubber and metal components, and the inter segment group, the activity term is positive and significant, indicating that in these cases countries with a stronger rule of law are more likely to form intra=firm trade ties. There is a contrast for the contract enforcement activity; for instance, with electrical parts there is a negative and significant effect. In this segment better, quicker contract enforcement is associated with a greater propensity to form intra-firm trade ties. In the case of engine production, the opposite is observed; weaker, slower contract enforcement is associated with the tendency to form intra-firm trade.

The difference effects also exhibit contrasts between rule of law and contract enforcement. For the rule of law, the electrical part and rubber and metal components segment there is a positive effect, this indicates that there is a tendency for intra-firm trade amongst countries with larger differences in the strength of the rule of law. For inter segments, there is a negative effect, indicating that intra-firm trade between segments of the automotive sector is more likely between countries with a similar level of rule of law. For contract enforcement, there is only a significant effect in engine, which indicates that intra-firm trade of engines is more likely in countries with similar contract enforcement levels.

The rule of law and contract enforcement results indicate mixed support for the property rights theory of intra-firm trade, where there are contrasting results between the two measures and across segments of the automotive supply chain.

The investment activity term is positive and significant in the three main groups (yet not in the inter segment group). This suggests that countries with a higher number of firms present and operating in the segment are associated with a propensity to form intra-firm trade ties, as observed for the three main segments. The investment difference is negative and significant in the three main segments, which indicates that countries with similar level of investment and firms present are more likely to be involved in intra-firm trade. This provides potential to inform on industrial policy regarding intra-firm trade, noting the relationship between firm investment and intra-firm trade within a segment of an important global supply chain (the automotive sector).

## 6. Robustness Checks

As part of the robustness checks for this study to ensure that the intra-firm configuration (as presented in fig 2) is a significant feature of the multilevel network for each component group, we employ the Z-Score approach outlined by Milo et al. (2002), and described in further detail in section 4.

If the Z-Score is above 2, this indicates that the intra-firm trade configuration occurs more in observed multilevel network than the simulated network. If the Z-Score is below -2 this would indicate that the intra-firm configuration (or motif) occurs less in the observed network than the simulated networks.

Fig 8 presents the Z-Score results, presenting the significance profile of the intra-firm trade configuration (or motif) for each component group. The results indicate that for the engines, rubber & metal parts and the inter-segment component groups, the intra firm trade configuration occurs significantly more in the observed networks than in the simulated networks. In other words, for these groups it is a significant feature of the multilevel data, especially in the cases of engines and rubber & metal parts. However, in the case of electrical parts, the Z-Score is between -2 and 2, indicating the number of intra-firm trade configurations in the electrical parts multilevel network is not significantly different from the simulated networks (generated under a random process). This suggests that caution should be applied when interpreting the electrical parts intra-firm trade network, as there is the potential for the over estimation of intra-firm trade ties in this component group.

***Figure 8 Z-Score Significance Profile Plot***

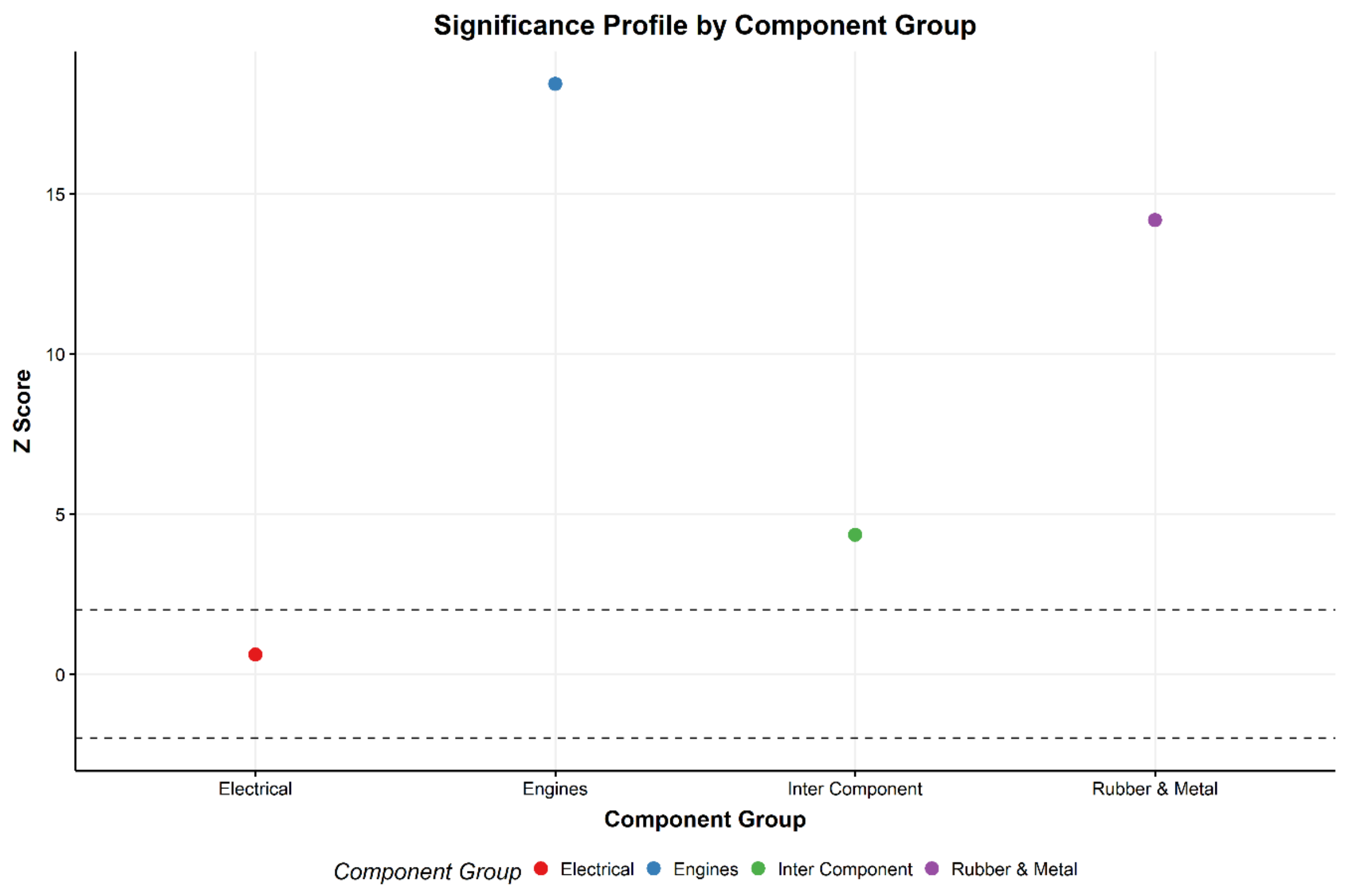

## 7. Discussion

This paper provides a novel approach to construct network of intra-firm trade amongst countries. This work contributes to existing literature utilising a multilevel perspective to understand economic networks, along with the stream of empirical research attempting to better understand the determinants of intra-firm trade. This paper addressed three central research questions through the use of an ERGM applied to unique network of intra-firm trade. The first research question asked whether intra-firm trade in various segments of the automotive supply chain adhered to the gravity model. However, the results indicate that there is mixed support for the gravity model; that intra-firm trade does not consistently adhere to the gravity model across segments of the automotive production chain. For instance, distance does not play a significant role, yet countries that share a border and common language are more likely to form intra-firm trade ties in the sector. The GDP results do not indicate that larger nations gravitate towards each other, rather point toward larger countries linking, via intra-firm trade, to many smaller nations.

The second research question asked whether the property rights theory explained patterns of intra-firm trade within the sector. Similar to the gravity model, there is mixed support that intra-firm trade adheres to the property rights theory, in particular given the contrasting results for contract enforcement and rule of law.

The final research question asked whether the patterns of intra-firm trade were consistent across various segment of the automotive supply chain. The initial network visualisations address this research question, where the network structure clearly differs, especially between electrical part and inter segment, with engines and rubber and metal parts. This was further emphasised by the ERGM results. The results indicate that there is a link between complexity of the product and transaction and patterns of intra-firm trade.

The ERGM also informed on the structural characteristics of intra-firm trade, which again highlighted the differences between segments of the automotive supply chain; for instance, there is only a strong tendency for clustering in the engines segment. The degree term indicates that in engines and rubber and metal parts, intra-firm trade ties are more hierarchical, concentrated in a small handful of actors, contrasting to electrical and inter segment, suggesting the spread of intra-firm trade depends on the technological content of the product group.

It is important to note, when discussing the ERGM results, that the robustness checks indicated that the intra-firm trade configuration was a significant feature of each component group (as shown by the Z-Score analysis), with the exception of the electrical parts groups. Therefore, caution is required when interpreting the ERGM results for electrical parts, as there is the potential for over-estimating intra-firm trade (i.e., the network may indicate an intra-firm trade tie that may, in reality, be absent).

There are several avenues for further research. What is highlighted by this analysis is that more is needed to develop theory to better understand intra-firm trade at the country level, and how this might differ by sector. Further work is required to develop datasets to inform on patterns of intra-firm trade. This work provides an initial attempt to address the calls by the UN Statistical Commission. Next steps to develop this line of research would be to develop multilevel datasets (that could be used for the construction of an intra-firm network. Where these datasets could be expanded in terms of the sectors A further area is the spread of shocks in the international trade network, examining how the spread differs between countries linked by potential intra-firms ties, to other forms of trade; as note that production network characterised by intra-firm trade can be more resilient in an economic downturn (Lanz and Miroudot 2011).

This study on intra-firm trade is subject to a number of limitations. Firstly, this work assumes that firms involved in the intra-firm configuration (or motif) used to filter and construct the intra-firm trade network engage in trade. However, the extant literature suggests that intra-firm is skewed and concentrated in a handful of firms (Ramondo et al. 2016; Alviarez and Saad 2023), therefore the intra-firm trade networks only represent potential intra-firm trade networks. In addition, there is the risk of over-estimating the number of intra-firm trade ties. In this study, we address this risk through the robustness checks implemented, to identify whether the intra-firm trade configuration is a significant feature of the multilevel data used to construct the intra-firm trade network. Given that intra-firm data is widely unavailable, this approach (with some limitations) provides a novel approach to enhance our understanding of intra-firm trade.

## 9. Appendix

Table 4 indicates the SITC codes used to create each component group for the three segments of the automotive production network.

***Table 4 Component groups and SITC codes***

| | |
|---|---|
| **Electrical Parts** | |
| SITC Rev. 3 Code | Code Description |
| 76211-76212 | Radio-broadcast receivers not capable of operating without an external source |
| 77812 | Electric accumulators (storage batteries) |
| 77823 | Sealed-beam lamp units |
| **Engines & Parts** | |
| SITC Rev. 3 Code | Code Description |
| 71321–71322 | Reciprocating internal combustion piston engines for propelling vehicles |
| 71323 | Compression ignition internal combustion piston engines (diesel or semi-diesel) |
| 77831 | Electrical ignition or starting equipment of a kind used for spark ignition |
| 77833 | Parts of the equipment of heading 778.31 |
| 77834 | Electrical lighting or signalling equipment |
| **Rubber & Metal Parts** | |
| SITC Rev. 3 Code | Code Description |
| 6251-62551 | Tyres, pneumatic, new, of a kind used on motor cars (including station wagon) |
| 62559 | Tyres, pneumatic, new, other |
| 62591 | Inner tubes |
| 62592 | Retreaded tyres |
| 62593 | Used pneumatic tyres |
| 62594 | Solid or cushion tyres, interchangeable tyre treads and tyre flaps |
| 69915 | Other mountings, fittings and similar articles suitable for motor vehicle |
| 69961 | Anchors, grapnels and parts thereof, of iron or steel |

Fig 9 to 12 provides the goodness of fit (GOF) plots for the ERGM models implemented in the paper. The GOF plots compare the salient structural features of the observed networks with a set of networks simulated from the estimated ERGM. If there is a good fit, then the simulated networks should share characteristics with the observed networks. The solid black line represents the features of the observed network, and the boxplots are the features of the simulated networks (based on the ERGM). We generally observe a good fit for the ERGMs.

*Figure 9 Electrical parts GOF*

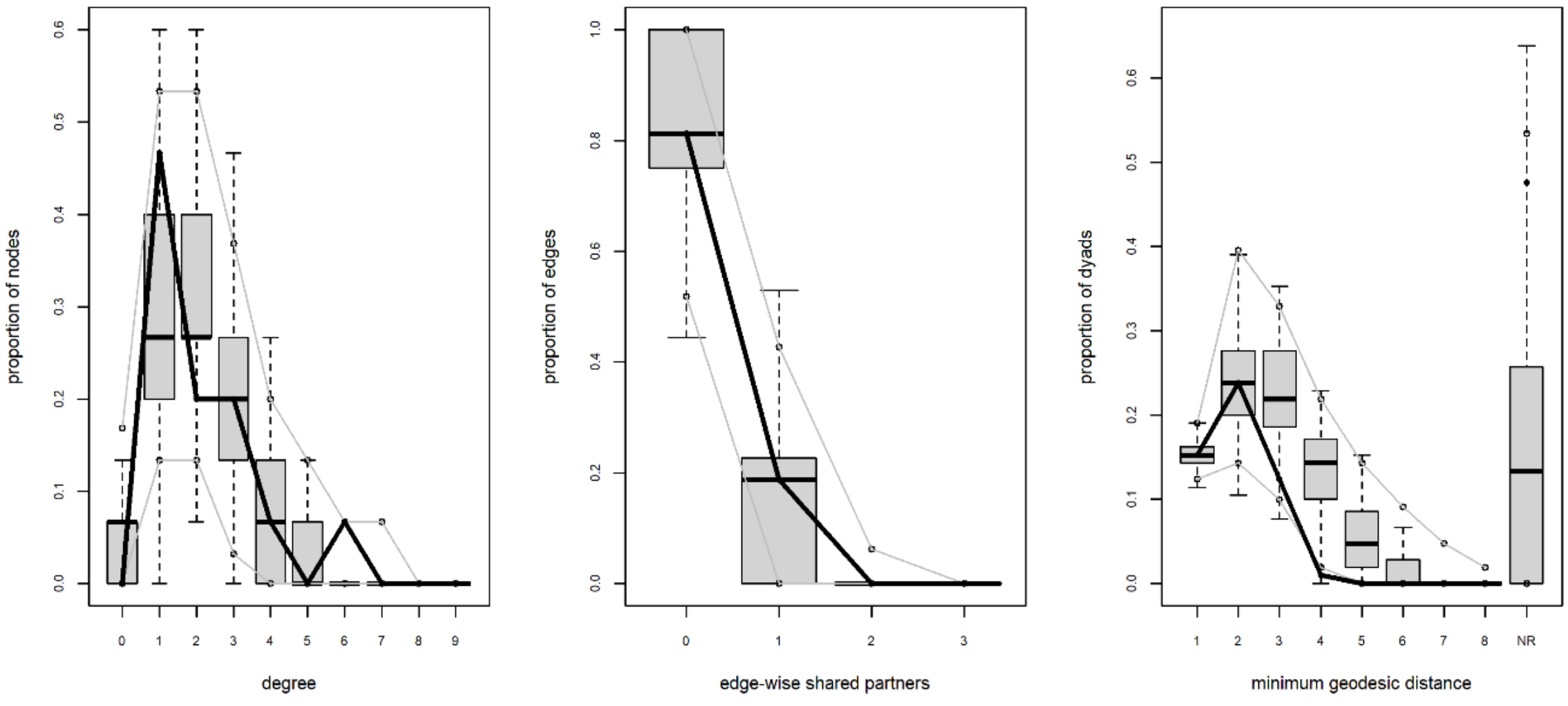

*Figure 10 Engines GOF*

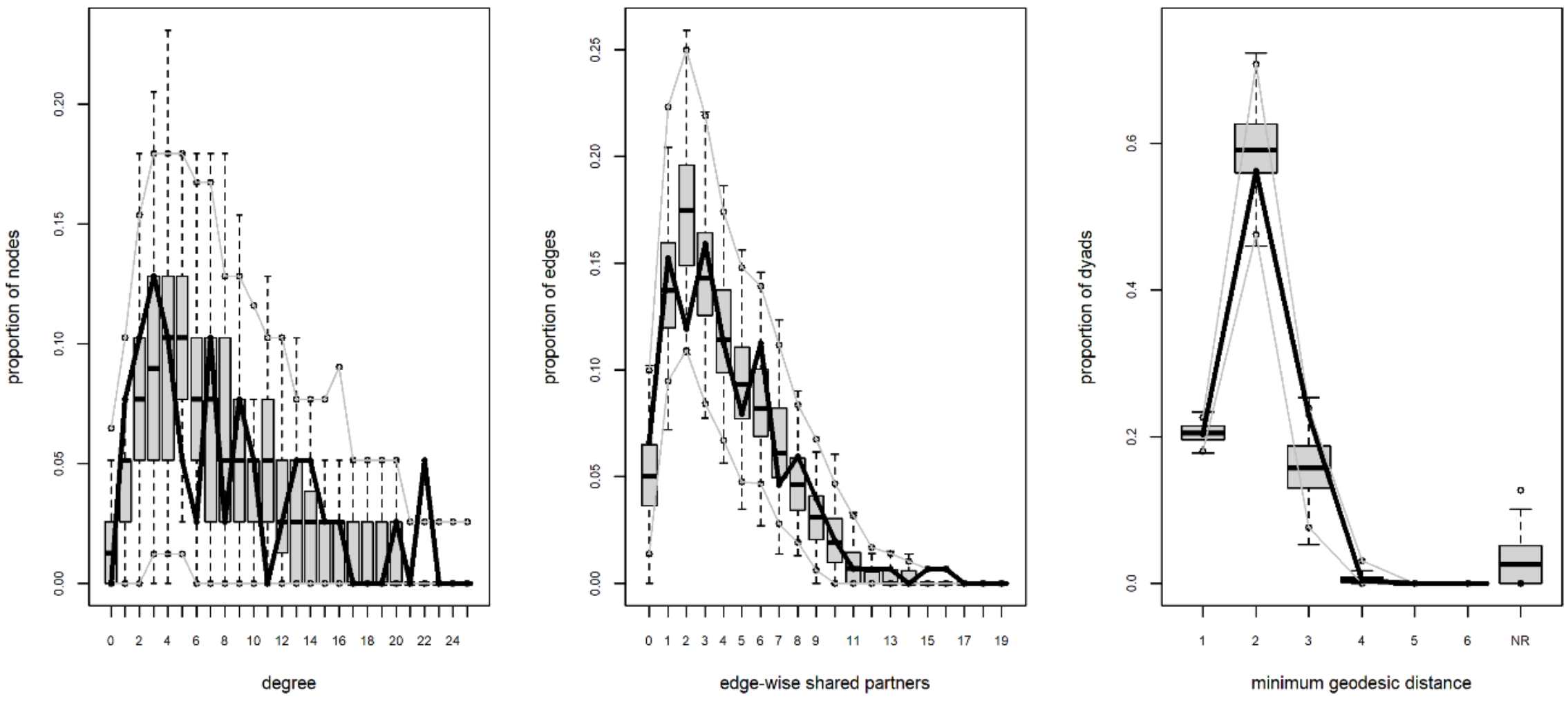

***Figure 11 Rubber & Metal GOF***

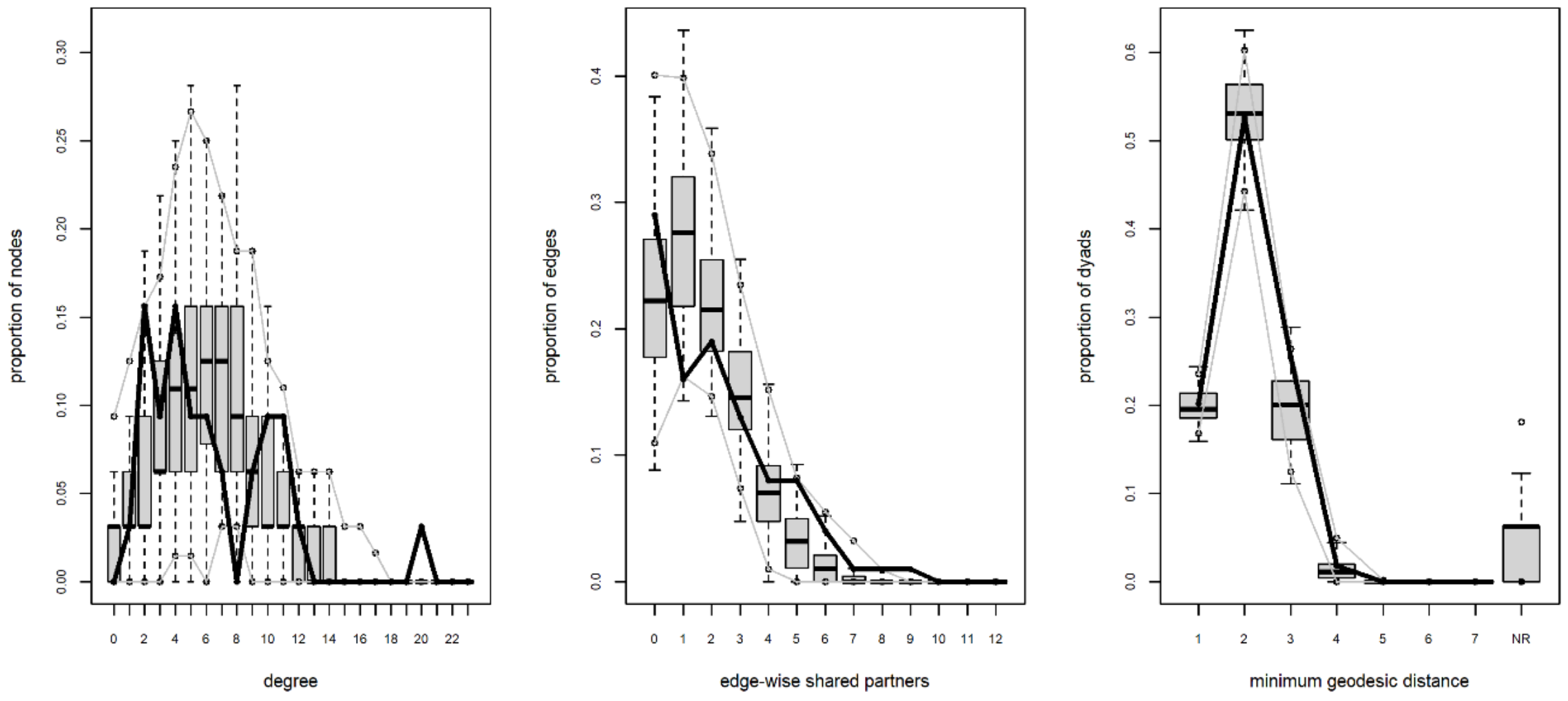

***Figure 12 Inter Segment GOF***

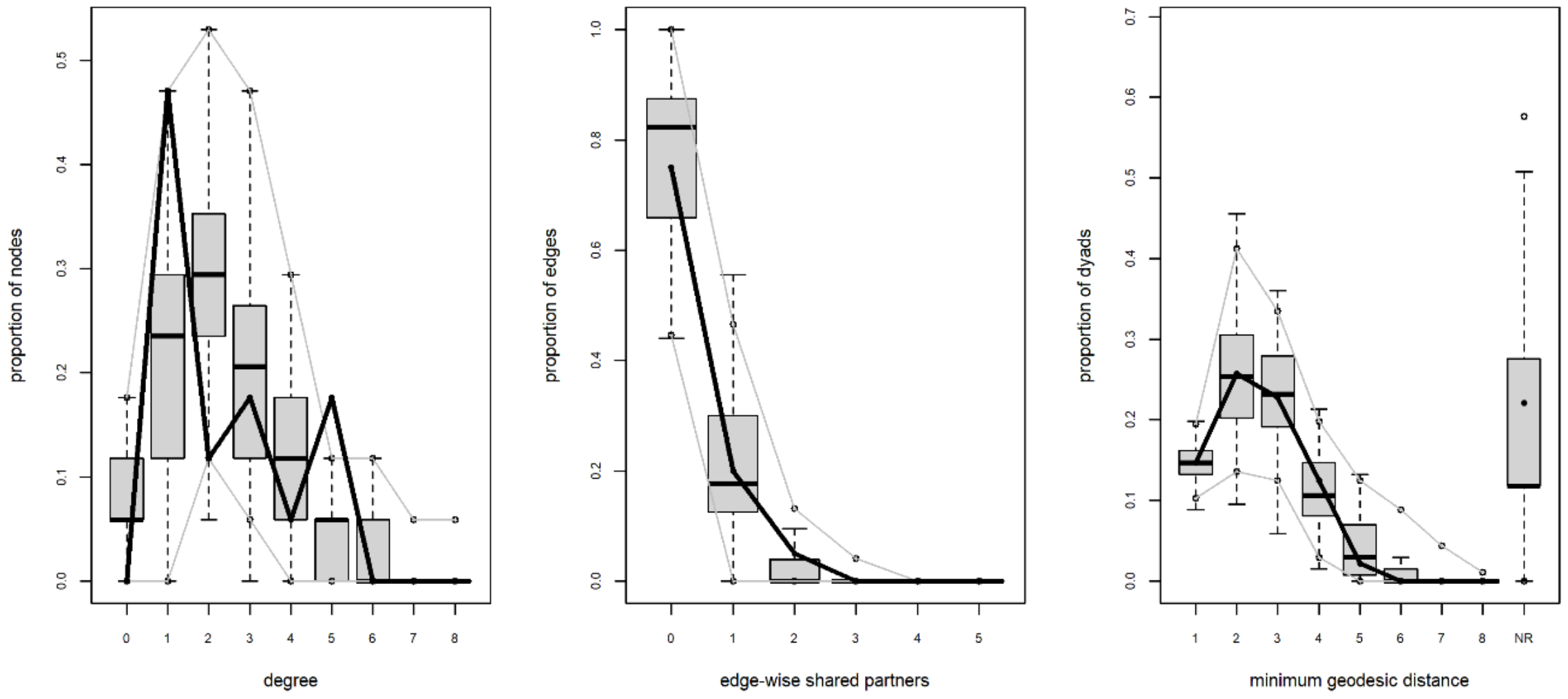